\documentclass[12pt]{iopart}
\usepackage{epsf}
\begin{document}

\title[Topological Quantum Gates with Quantum Dots]
{Topological Quantum Gates with Quantum Dots}

\author{Jiannis K. Pachos\footnote[3]{jiannis.pachos@imperial.ac.uk} and Vlatko
Vedral\footnote[1]{v.vedral@imperial.ac.uk}}

\address{ Blackett Laboratory, Imperial College London, Prince Consort Road,\\
London, SW7 2BW, UK}

\begin{abstract}

We present an idealized model involving interacting quantum dots that can support
both the dynamical and geometrical forms of quantum computation. We show that by
employing a structure similar to the one used in the Aharonov-Bohm effect we can
construct a topological two-qubit phase-gate that is to a large degree independent
of the exact values of the control parameters and therefore resilient to control
errors. The main components of the setup are realizable with present technology.

\end{abstract}


\submitto{\JOB (special issue on Quantum Computing)}

\maketitle

\section{Introduction}

A very promising scenario for the implementation of quantum computation is
considered to be the solid state realization of quantum dots
\cite{Divincenzo1,Divincenzo2,Briggs}. The solid state arena is attractive because
there is a great deal of technological knowledge already accumulated from the domain
of classical computation. However, if we are to use solid state devices for quantum
computation, we need to manipulate individual quantum systems, like
electrons in quantum dots, that demand a much higher degree of control accuracy than
currently available. There has been a number of recent proposals to address the
issue of controlability by using geometrical and topological effects
\cite{Kitaev,Jiannis}. The key advantage of these methods is that the resulting
geometrical and topological gates do not depend on the overall time of the
evolution, nor on small deformations in the control parameters. Possible
manifestations of geometrical phases are the Berry phases obtained, for example,
through a cyclic adiabatic evolution of a system
\cite{Berry} or through the Aharonov-Bohm effect \cite{AB}. Within solid state physics,
there have even been proposals to implement Berry phases with Josephson Junctions
\cite{Vlatko} as well as Aharonov-Bohm phases encoded in the different spin states
of electrons manipulated in quantum dot structures \cite{Akera,Loss}.

In this paper we present a simple solid state implementation of a charge based
structure that is capable of supporting both dynamical and geometrical quantum
computation. Consider quantum dots that can either be empty or can accommodate an
electron. In an array of quantum dots we assume that we are able to lower the
potential between any two dots and facilitate the quantum tunneling between them.
Intrinsically, the Hamiltonian of this system is governed mainly by three parts,
namely the potential wall of height $V_W$ separating two successive dots, the
Coulomb interaction $V_C=e^2/r$ between two electrons which is active when they are
occupying the same dot or adjacent dots in a close proximity and the external laser
fields which, in principle, can facilitate to move an electron along specific paths.

\begin{figure}[h]
\begin{center}
\hspace{1cm}
\epsfbox{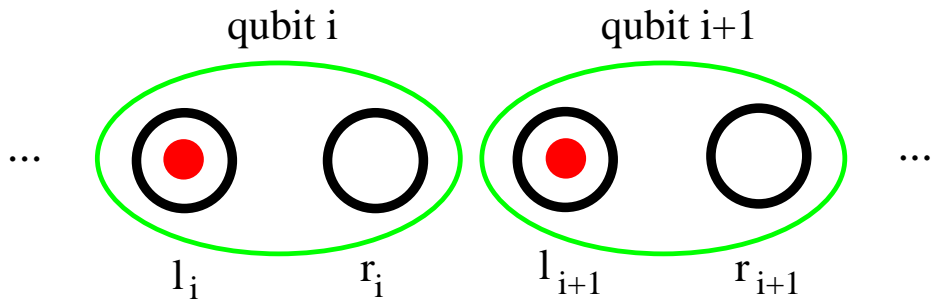}
\end{center}
\caption{\label{array}The array of the dots that comprises an array of qubits.
Each qubit is represented by two dots and one electron. The electron can be in the
quantum state of occupancy superposition between the left and the right dot
controlled by a tunneling procedure that produces a one qubit rotation.}
\end{figure}

Imagine that we have an array of paired dots and that to each pair corresponds an
electron as in Fig. \ref{array}. We can interpret the qubit $i$ to be encoded by the
pair of dots $i$; if the electron is in the left dot ($l_i$) it corresponds to the
logical state $|0\rangle$, while when it is in the right dot ($r_i$) this
corresponds to the state $|1\rangle$. Preparation of the initial qubit state as well
as the final read out are technically easy tasks. The performance of one qubit gates
is achieved by lowering the potential wall $V_W$ between the pair of the dots and,
hence, quantum tunneling between the two dots will create superpositions of the
logical states \cite{Bonadeo} giving, for example, the state $|\psi\rangle=c_0
|0\rangle + c_1 |1\rangle$. Thus, any possible one qubit rotation can in principle
be performed.

For a two-qubit phase-gate we need to perform a controlled transition where the
state of one of the qubits is changed conditional on the state of the second qubit.
We shall present in Section \ref{gate} three different ways of performing such a
gate in a dynamical, geometrical and topological fashion. A controlled phase-gate
can then allow us to execute any arbitrary quantum computation providing that we can
also implement single qubit gates. Finally, in Section \ref{impl}, a set of
different potential implementations are presented based on current solid state
technology.

\section{Two qubit phase-gates}
\label{gate}

As we have seen the charge based quantum computation model consists of two dots $l$
and $r$ and an electron which can be in a state that is a superposition of occupying
both of them. Indeed, if the electron is in the left dot then the state of the
system can be represented as $|n_l=1,n_r=0\rangle$, where $n_k$ is the occupation
number of dot $k$, while if the electron is in the right dot then the state is
written as $|n_l=0,n_r=1\rangle$. These states correspond to the logical $|0\rangle$
and $|1\rangle$ qubit states. How such a system can support entangling gates between
the qubits is described in the following.

\subsection{Dynamical phase-gate}
\label{dyn}

In the purely dynamical implementation of the controlled phase-gate we rely on the
fact that when two electrons are trapped in the neighboring wells, then their
dynamical phase due to the Coulomb interaction is larger than when the electrons are
far apart. Hence, when the dots are in the state $|01\rangle$, their extra phase
with respect to all the other states is equal to $e^{i\Delta E t}$, where $\Delta E$
is the increase in the electron energy due to the Coulomb repulsion and $t$ is the
time during which the electrons are close enough to exhibit a non-negligible
interaction. The evolution of the system of the two qubits in the basis
$|00\rangle$, $|01\rangle$, $|10\rangle$ and $|11\rangle$ is then given by
\[
U_{dyn}= \left( \begin{array}{cccc} 1 & 0 & 0 & 0
\\
0 & e^{i \Delta E t} & 0 & 0
\\
0 & 0 & 1 & 0
\\
0 & 0 & 0 & 1
       \end{array} \right) .
\]
After time $t = \pi / \Delta E$ we obtain a dynamical version of the two-qubit
phase-gate. This gate is, in fact, algorithmically equivalent to the controlled-not
gate and is capable of generating entanglement between two initially disentangled
qubits. Most of the current proposals for implementation of quantum computation are
based on dynamical gates.

\subsection{Geometrical phase-gate}

Now we would like to show how to implement the same gate in the same setting, but
using the geometrical instead of the dynamical phase. Consider a homogeneous
magnetic field $B$ in the neighborhood of the dots that comprise the logical array
of qubits. An electron which spans a closed trajectory (loop) inside the magnetic
field will acquire a phase factor proportional to the flux of the magnetic field
encircled by the loop \cite{AB}. The electron can be moved around by applying an
additional external time dependent laser field that guides the electron along the
desired trajectory on a plane perpendicular to the magnetic field $B$.
\begin{figure}[h]
\begin{center}
\hspace{1cm}
\epsfbox{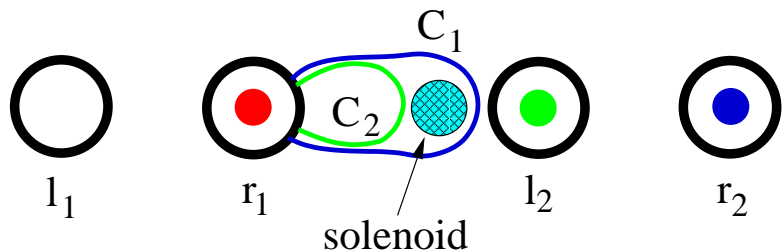}
\end{center}
\caption{\label{loops}The model for geometrical two qubit gate. If the
electron of the second pair is in the right position then the electron in the first
pair traverses the loop $C_1$, while if it is in the left position the traversed
loop is $C_2$.}
\end{figure}
In the case of the controlled interaction we want to achieve, we can imagine that we
encourage a cyclic evolution of an electron from the dot $r_1$ towards the dot $l_2$
of the second pair of dots as seen in Fig. \ref{loops}. If there is no electron in
the dot $l_2$, then the loop $C_1$ of the first electron spans a surface given by
$S_1$ and it is determined only by our controlled procedure since the electrons are
non-interacting. As a result, after the first electron has returned to dot $r_1$, it
has acquired a phase given by $\phi_1=\int\int_{S_1} B $. On the other hand, if
there is an electron in dot $l_2$ then the trajectory of the first electron will be
influenced by the additional Coulomb interaction (repulsion) and it will span a loop
$C_2$ that encloses a smaller surface area $S_2$. Hence, when the electron has
returned to dot $r_1$, then it has acquired a phase given by $\phi_2=\int\int_{S_2}
B \neq \phi_1 $. It is clear that no non-trivial evolution will occur if the
electron of the first qubit is in dot $l_1$. The unitary evolution that is finally
implemented in this way is of the form
\[
U_{geom}=
\left( \begin{array}{cccc}
e^{i \phi_1} & 0 & 0 & 0
\\
0 & e^{i \phi_2} & 0 & 0
\\
0 & 0 & 1 & 0
\\
0 & 0 & 0 & 1
       \end{array} \right) ,
\]
which, up to a local phase rotation of qubit $2$, is equivalent to a controlled
phase-gate that changes only the state $|00\rangle$ to the state
$e^{i(\phi_1-\phi_2)}|00\rangle$. This is the geometrical version of a two qubit
gate which was previously implemented dynamically.

\subsection{Topological phase-gate}

We notice that if there is a small change in the geometry of the loop in the
previous evolution, then the form of the phase-gate will also change. This is a
drawback of the previous model as, in general, the external field guiding the
electron will fluctuate and therefore the acquired phase, which is proportional to
the encircled magnetic flux due to the motion of the electron, will also suffer from
this error in our control. We can remedy this problem by spatially restricting the
magnetic field in the Aharonov-Bohm effect. The gate we now describe will be
independent of the actual shape of the electron trajectory to a high degree. To
achieve this, we need to place between each pair of dots a small solenoid with a
large density magnetic field confined inside it \cite{Chopra} as depicted in Fig.
\ref{loops}. The control procedure of the gate is the same as the one presented in the
geometrical case, only that now the solenoid is positioned inside the surface area
$S_1-S_2$. This is the surface which is not covered by the electron of qubit $1$ if
and only if the electron of qubit $2$ is in the dot $l_2$. The net effect resulting
from this procedure is an evolution in the computational basis of the two qubits
given by
\[
U_{top}=
\left( \begin{array}{cccc}
1 & 0 & 0 & 0
\\
0 & e^{i \Phi} & 0 & 0
\\
0 & 0 & 1 & 0
\\
0 & 0 & 0 & 1
       \end{array} \right) ,
\]
where $\Phi$ is proportional to the magnetic flux confined by the solenoid. In other
words, by employing the Coulomb interaction it is possible for electron $1$ to
acquire a fixed phase $\Phi$ if it circulates around the solenoid, while no phase is
obtained if the electron does not circulate around it. The phase $\Phi$ is
independent of the exact shape of the electron's trajectory. This is, therefore, the
topological version of the quantum phase-gate discussed before. Unlike the previous
gates, it has the advantage that any deformation of the trajectory of the electron
that does not alter its winding number around the solenoid has no effect whatsoever
on the gate. Consequently, this offers us a high degree of independence from control
errors.

\section{Solid state implementation}
\label{impl}

The realization of the topological gate requires the confinement of the magnetic
field to the very small area of the solenoid which is placed in-between the dots. We
would like now to estimate how small the radius of the solenoid should be in order
to compare it with the experimental state of the art. For simplicity we assume that
the electron is confined in a one dimensional harmonic potential $V_h= m_e \omega^2
x^2 /2$, where $m_e$ is the electron mass, $\omega$ is the trapping frequency and
$x$ is the displacement of the electron. Consider the case where we want to move an
electron from dot $r_1$ (see Fig. \ref{loops}) towards dot $l_2$ positioned at
$x=0$, by displacing the trapping potential. If there is no electron in $l_2$ then
the trapping potential carrying the electron from dot $r_1$ moves so that its final
minimum is positioned at $x=0$. If the electron exists at $x=0$, then the final
position of the traveling electron is determined by the minimum of the combined
trapping potential and the Coulomb repulsion. An approximate estimate of the
displacement of the new minimum is given by
\begin{equation}
\Delta x \approx{10 \over \omega ^{2/3}} .
\end{equation}
Typical trapping frequencies are in-between $10^6 Hz$ and $10^9 Hz$ which results in
$\Delta x$ in-between $1mm$ and $100\mu m$. This distance is roughly the size of the
region in which the magnetic field has to be confined in order to implement the
topological evolution. The present technology \cite{Chopra} allows us to construct a
solenoid with the dimensions smaller than $1\mu m$, which is more than sufficient
for the implementation of our proposal.

So far in our proposal the electron from dot $r_1$ has to continuously move from its
original position to $l_2$ close to the other electron and then back to $r_1$.
However this does not need to be continuous as the electron can, in fact, perform
the same looping trajectory by discrete tunneling between, for example three
different dots.
\begin{figure}[h]
\begin{center}
\hspace{1cm}
\epsfbox{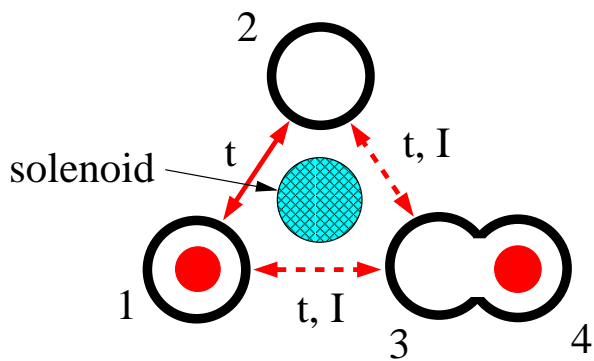}
\end{center}
\caption{\label{qdot1}Quantum dot implementation of the traversing loop by tunneling
transitions. The double dot allows for the presence of two well distinguished
electrons in each of its sides which exhibit Coulomb repulsion with potential $U$.}
\end{figure}
Towards that direction, let us consider the implementation of the two qubit
phase-gate as presented in the previous analysis, but with the configuration of four
dots arranged as in Fig. \ref{qdot1}. Consider a magnetic field $B$, that is large
enough so that the spins in the dots are aligned along the magnetic field, and
henceforth play no role in the subsequent evolution. Assume a double dot system
where $U$ is the Coulomb repulsion with single occupancy of each dot in the double
dot system while at the same time no tunneling occurs between them. Allowing in
addition the tunneling process to take place between neighboring dots, the resulting
Hamiltonian of the system is given by
\begin{equation}
H=H_d+H_C+H_t
\end{equation}
where
\begin{equation}
H_d=\sum_i E_{i} d_i^\dagger d_i \,\,\, ,\,\,\,\,\, H_C= Ud_3^\dagger d_3
d_4^\dagger d_4 \,\,\, ,\,\,\,\,\, H_t= \sum _{\langle i,j \rangle} (t_{ij}
d_i^\dagger d_j +H.c.)
\end{equation}
where $d_i$ is the annihilation operator of the electrons in the dot $i$, $E_i$ is
its free energy, $t_{ij}$ is the tunneling coupling between the neighboring dots $i$
and $j$ and ${\langle i,j \rangle}$ indicates nearest neighbors. Assume that the
tunneling couplings $t_{ij}$ can be turned on and off at will, obtaining the maximum
value $t_{12}=t_{23}=t_{31}=J$, and we shall assume for simplicity that $U$ is also
the Coulomb potential for two electrons within the same dot. If dot $4$ is occupied,
then the effective tunneling between $2$ and $3$ or $1$ and $3$ is $I=2 J^2/U$. We
shall consider the limit of large $U$ where we are in the Coulomb blockade regime,
i.e. $I\ll J$ and hence, in that case, actual tunneling does not occur. In that case
it is impossible to circulate an electron initially in the dot $1$ around the dots
$1$, $2$, $3$ and back to $1$ if there is an electron in the dot $4$, by turning on
successively the couplings $t_{12}$, $t_{23}$ and $t_{31}$ for time $T=\pi/(2J)$ to
transit the electrons from one dot to the other. On the other hand, if the dot is
empty, an electron can go around the closed path acquiring a phase due to the
Aharonov-Bohm effect. The latter is given by $AB$ where $A$ is the area of the
triangle spanned by the dots $1$, $2$ and $3$. In order to guarantee that the
electron initially in $1$ will return back to its initial position if there is an
electron in $4$ we have to activate $t_{12}$, $t_{23}$ then again $t_{12}$ (the
electron returns back in the case it is still in $2$ due to Coulomb blockade) and
then $t_{31}$. While $t_{31}$ is activated, the electron in $1$ does not move again
due to the Coulomb blockade. Hence, at the end of this evolution the system returns
back to the qubit states where there is no occupancy in the dots $2$ or $3$. The
resulting two-qubit gate is exactly the same as given in the previous section with
the phase given by $\phi= A B e/h$. Alternatively, the magnetic field can be
confined in a solenoid imbedded in-between the dots $1$, $2$ and $3$. Therefore,
this simple and realistic systems offers a possibility to implement the topological
based quantum computation discussed at a more abstract level in Section $2$.

\section*{Discussion and Conclusions}

We have presented here three possible implementation of quantum gates. Every
subsequent implementation, although more difficult to implement, offers a higher
degree of reliability with respect to control errors. In the topological
implementation, if the magnetic field is confined to a very small region of space
between the quantum dots, we can achieve an arbitrary high fidelity of gate
implementation. We even saw from a very simple analysis in the previous section that
this is, in principle, possible with current technology. However, our analysis does
not take into account other possible sources of errors, such as decoherence due to
the coupling of the electron with the environment. The environment can, for example,
act as a projective measurement determining the position of the electron thereby
destroying any superposition of the electron occupancy states of different dots.
Alternatively, the environment can destroy the phase coherence between different
element of the superposition. For the particular analysis of the behaviour of
geometric phases under classical and quantum noises see, for example,
\cite{Vourdas}. To successfully compensate this kind of errors, in parallel to the
methods presented here, we will, most likely, have to resort to other existing
methods like quantum error correcting codes \cite{Steane} or error avoiding methods
like in decoherence-free subspaces \cite{Almut}. Note finally that we can interpret
our structure as generating a single electron circulating current (vortex) by the
electron in dot $r_1$ conditional on the presence of an electron in $l_2$. This is a
very interesting physical system in its own right that could be potentially used for
other applications of quantum circuits, for example in measuring the flux of the
magnetic field by the resulting relative phase between the two distinct possible
evolutions. A similar model with three potential wells has also been used to
describe generation of vortices in trapped Bose-condensates \cite{Milburn}. An
elaborated analysis of our proposal including various decoherence mechanisms is
therefore very much worthwhile and will be presented elsewhere.

\ack

The authors would like to thank A. Briggs for pointing out reference \cite{Chopra}.
This work was supported in part by the European Union, the U.K. Engineering and
Physical Sciences Research Council and Elsag Spa.

\end{document}